\documentclass{jpsj3}
\title{Pseudogap, Superconducting Gap, and Fermi Arc in High-$T_c$ Cuprates Revealed by Angle-Resolved Photoemission Spectroscopy}

\author{Teppei Yoshida$^1$\thanks{E-mail address: yoshida@wyvern.phys.s.u-tokyo.ac.jp}, Makoto Hashimoto$^2$, Inna M. Vishik$^2$, Zhi-Xun Shen$^2$, Atsushi Fujimori$^1$}

\inst{$^1$Department of Physics, University of Tokyo, Bunkyo-ku,
Tokyo 113-0033, Japan\\ $^2$Stanford Institute for Materials and
Energy Sciences, SLAC National Accelerator Laboratory, 2575 Sand
Hill Road, Menlo Park, CA 94025, USA }

\abst{We present an overview of angle-resolved photoemission
spectroscopy (ARPES) studies of high-temperature cuprate
superconductors aiming at elucidating the relationship between the
superconductivity, the pseudogap, and the Fermi arc. ARPES studies
of underdoped samples show a momentum dependence of the energy gap
below $T_c$ which deviates from a simple $d$-wave form, suggesting
the coexistence of multiple energy scales in the superconducting
state. Hence, two distinct energy scales have been introduced,
namely, the gap near the node (characterized by $\Delta_0$) and in
the anti-nodal region (characterized by $\Delta^*$). Dichotomy
between them has been demonstrated from the material, doping, and
temperature dependence of the energy gap. While $\Delta^*$ at the
same doping level is approximately material independent,
$\Delta_0$ shows a strong material dependence tracking the
magnitude of $T_{c,max}$. The anti-nodal gap does not close at
$T_c$ in contrast to the gap near the node which follows something
closer to a BCS-like temperature dependence. An effective
superconducting gap $\Delta_{sc}$ defined at the end point of the
Fermi arc is found to be proportional to $T_c$'s in various
materials.}

\kword{high-$T_c$ superconductor, angle-resolved photoemission
spectroscopy}

\begin{document}
\maketitle

\section{Introduction}
%Preformed Cooper pair vs Competing order
From the earliest studies of high-$T_c$ cuprate superconductors,
the origin of the pseudogap has remained unresolved and has been
considered as the most fundamental problem for understanding the
mechanism of high-$T_c$ superconductivity. The central issue is
whether the pseudogap is related to the superconductivity or is a
distinct phenomenon from the superconductivity. In the former
scenario, a possible origin of the pseudogap is preformed Cooper
pairs lacking phase coherence \cite{Anderson, Kivelson}. In that
picture the pseudogap is a remnant of the superconducting gap into
the phase-incoherent regime and reflects the intrinsic pairing
tendency of the underlying electronic states. In the latter
scenario, the pseudogap arises from a competing order or
fluctuations such as a density-wave \cite{DDW,Hanaguri,Wise} or
antiferromagnetic correlation \cite{Kamimura,Prelovsek}.

\begin{figure}
\begin{center}
\includegraphics[width=8cm]{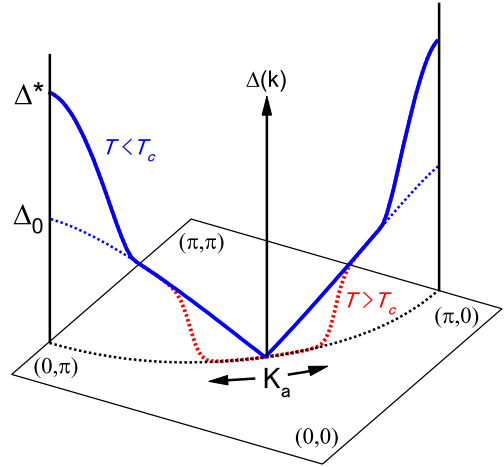}
\end{center}
\caption{(Color online) Schematic picture of the energy gap on the
Fermi surface of the high-$T_c$ cuprate superconductor. Deviation
of the energy gap from the simple $d$-wave is characterized by the
$d$-wave-like nodal gap ($\Delta_0$) and the larger anti-nodal gap
($\Delta^*$). Above $T_c$, the gap in the nodal direction closes.
The Fermi arc length $K_a$ is defined by the zero gap region on
the Fermi surface slightly above $T_c$. The gap becomes simple
$d$-wave-like when $\Delta^*$ becomes as small as $\Delta_0$.}
\label{EnergyGap}
\end{figure}

%Two gap vs One gap
It has been well known for more than 15 years that the pseudogap
in the antinodal $\sim$($\pi$,0) region increases with underdoping
and finally becomes much larger than the BCS value of $2\Delta
\sim 4 k_BT_c$ as observed by angle-resolved photoemission
spectroscopy (ARPES) \cite{Campuzano} and tunneling spectroscopy
\cite{Miyakawa}. Such a large anti-nodal gap in the underdoped
materials had sometimes been attributed to strong Cooper pairing.
However, the energy gap measured by Andreev reflection
\cite{Deustcher}, temperature-dependence of penetration depth
\cite{Panagopoulos}, and Raman scattering of B$_{2g}$ geometry
\cite{Opel,Tacon}, the latter of which probes the gap near the
$d$-wave node and the former of which should be directly
associated with superconductivity, exhibit different trends. That
is, with underdoping, the gap magnitude near the node saturates
around the optimal doping and then decreases or remains nearly
constant, suggesting a different origin of the gap near the node
from the anti-node. Stimulated by those experimental results,
detailed investigations of the energy gap in the momentum space by
ARPES have recently been performed extensively. An ARPES study of
Bi$_2$Sr$_2$CaCu$_2$O$_8$ (Bi2212) has revealed the presence of
two distinct energy scales for the nodal and anti-nodal regions in
the heavily underdoped regime \cite{Tanaka}. A similar ``two-gap"
behavior has been observed in the single-layer cuprates for under-
and optimally-doped regions such as La$_{2-x}$Sr$_x$CuO$_4$ (LSCO)
and Bi$_2$Sr$_2$CuO$_{6+\delta}$ (Bi2201)
\cite{KondoPRL,HashimotoDisorder,YoshidaTwoGap,Terashima}. Also, a
temperature-dependent angle-integrated photoemission study of LSCO
has indicated two distinct energy and temperature scales of the
gap \cite{Hashimoto}. These observations are schematically
illustrated in Fig. \ref{EnergyGap} as a two-gap picture, where a
gap from competing order coexists with a superconducting gap. In
this picture, the gap in the anti-node $\Delta^*$ is sufficiently
larger than the gap near the node $\Delta_0$, and the gap function
shows a deviation from a simple $d$-wave form, indicating a
crossover between the two kinds of gaps. When $\Delta^*$ is
comparable to or smaller than $\Delta_0$, the clear deviation may
not exist. The existence of competing order, which has spatial
modulations of the charge like checkerboard, has also been
suggested by scanning tunneling spectroscopy studies (STM) on
Ca$_{2-x}$Na$_x$CuO$_2$Cl$_2$ \cite{Hanaguri}. Furthermore, NMR
studies have revealed two distinct temperature scales in the
relaxation rate corresponding to the pseudogap temperature $T^*$
and $T_c$ \cite{Zheng, Kawasaki}.

%On the other hand, attempts have also been made to understand the
%pseudogap phenomena within the single $d$-wave energy gap in other
%ARPES studies \cite{NormanArc,Valla,Kanigel,Meng,Shi}. In such a
%single gap picture of the pseudogap phenomena, preformed Cooper
%pairs are considered as the origin of the pseudogap. Thus, the
%question of whether the energy gap of the high-$T_c$
%superconductors can be explained by the two-gap picture or the
%one-gap picture has not been settled yet.

In this review article, we shall give an overview of the results
of ARPES studies dealing with the issue of the two gaps
\textit{versus} one gap. This article is written as follows:
First, we demonstrate the material- and doping-dependence of the
energy gaps. Next, we describe the temperature dependence of the
energy gaps. Finally, we discuss the relationship between the
observed energy gaps and $T_c$'s based on the Fermi arc picture
\cite{Oda,LeeWen}.

\section{Material and Doping Dependences of the Energy Gap}
It has been well known that $T_c$ in the optimally doped region
($T_{c,\textrm{max}}$) generally increases with the number ($n$)
of adjacent CuO$_2$ planes from single layer ($n=1$), double layer
($n=2$), to triple layer ($n=3$). However, it has been unclear
what governs the $n$ dependence of $T_{c,\textrm{max}}$. In this
section, we shall explain the energy gap of the single layer,
double layer, and triple layer cuprates, focusing on the deviation
of the gap function in the antinodal region from the simple
$d$-wave form.

\begin{figure}
\begin{center}
\includegraphics[width=16cm]{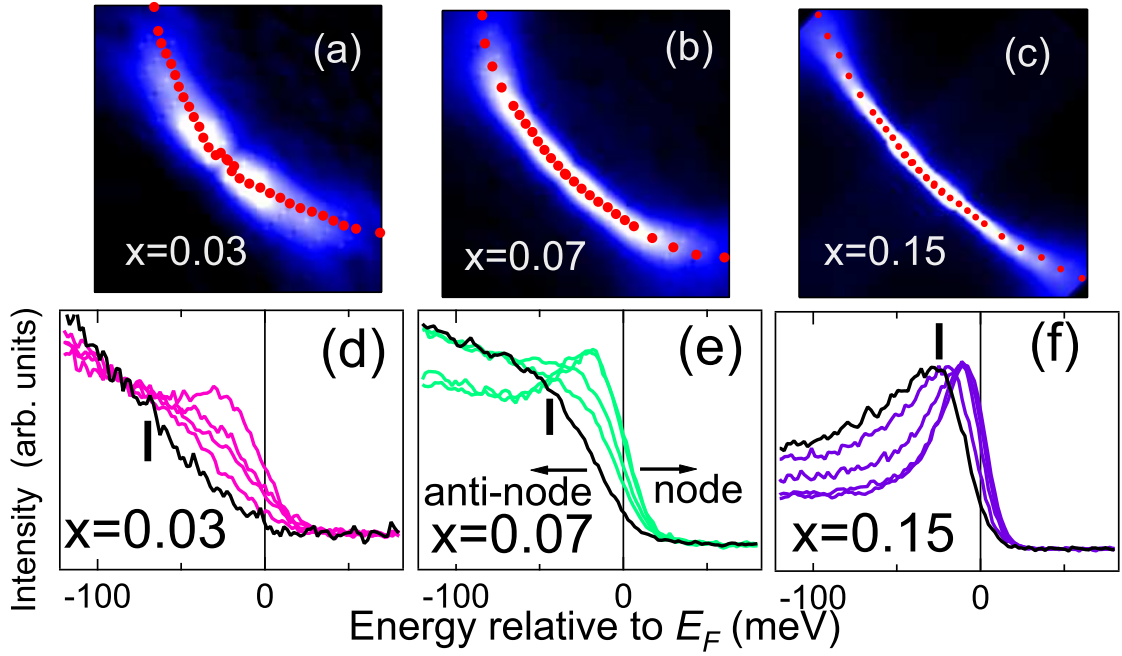}
\end{center}
\caption{(Color online) Fermi surface and energy distribution
curves (EDCs) at $k_F$ in La$_{2-x}$Sr$_x$CuO$_4$ (LSCO) with
various doping levels taken at $T$= 20 K \cite{YoshidaTwoGap}.
(a)-(c): Spectral weight mapping at $E_F$ in momentum space for
each doping level. Red dots indicate Fermi momenta $k_F$ defined
by the peak positions of the momentum distribution curves.
(d)-(f): EDCs at $k_F$ for each doping level. Black lines
correspond to antinodal EDCs and vertical bars represent energy
position of the antinodal gap.} \label{Yoshida1}
\end{figure}

\begin{figure}
\begin{center}
\includegraphics[width=16cm]{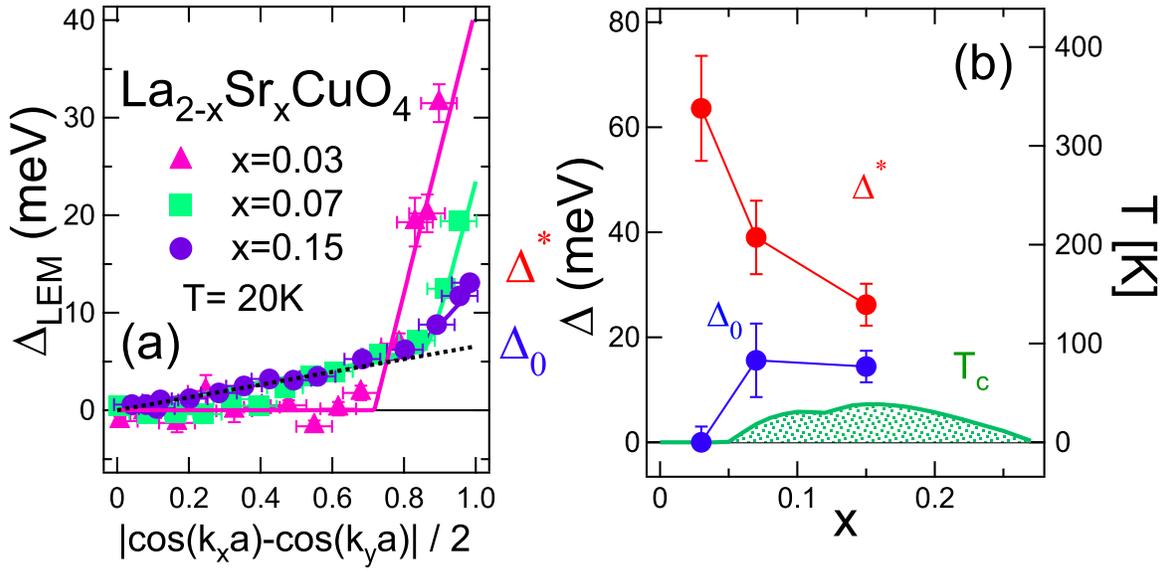}
\end{center}
\caption{(Color online) Momentum dependence of the energy gap at
$T$= 20 K in La$_{2-x}$Sr$_x$CuO$_4$ with various doping levels
\cite{YoshidaTwoGap}. (a): Leading edge midpoints (LEM)
$\Delta_\mathrm{LEM}$ relative to that at the node. (b): Doping
dependence of $\Delta^*$ and $\Delta_0$ obtained by assuming the
relationship $\Delta_\mathrm{peak} \simeq 2.2
\Delta_\mathrm{LEM}$.} \label{Yoshida2}
\end{figure}

%Single layer
First, we present the observation of the energy gap in the
single-layer high-$T_c$ cuprates La$_{2-x}$Sr$_x$CuO$_4$ (LSCO) to
define the two energy scales. In Fig. \ref{Yoshida1}, we show the
result of ARPES studies of LSCO by Yoshida \textit{et al.}
\cite{YoshidaTwoGap}. Panels (a)-(c) show spectral weight mapping
at $E_F$ taken at $T$=20 K for each doping level and clearly
illustrate the intensity suppression in the anti-node region due
to the gap being maximum there. Panels (d)-(f) show that the
leading edge midpoints (LEMs) of energy distribution curves (EDCs)
at the Fermi momenta $k_F$ shift toward higher binding energies in
going from the node to the anti-node, indicating an anisotropic
low-temperature gap. The angular dependence of the gap for each
doping is plotted as a function of the $d$-wave order parameter
$|\cos(k_xa)-\cos(k_ya)|/2$ in Fig. \ref{Yoshida2}(a). Near the
node (near $|\cos(k_xa)-\cos(k_ya)|/2 \sim 0$), this plot obeys a
straight line expected for the simple $d$-wave order parameter,
but it starts to deviate upward around
$|\cos(k_xa)-\cos(k_ya)|/2\sim$0.7-0.9. Nearly the same results
have been obtained for the optimally doped single-layer cuprates
Bi2201 \cite{KondoPRL,KondoNat}. In order to characterize these
behaviors of the energy gaps, one can define two energy scale
parameters $\Delta^*$ and $\Delta_0$: $\Delta^*$ is a gap closest
to $|\cos(k_xa)-\cos(k_ya)|/2=1$ and $\Delta_0$ is the
extrapolated value of the linear simple $d$-wave gap near the node
to $|\cos(k_xa)-\cos(k_ya)|/2=1$, as indicated in Fig.
\ref{EnergyGap} and Fig. \ref{Yoshida2}(a). In Fig.
\ref{Yoshida2}(b), the doping dependences of the $\Delta^*$ and
$\Delta_0$ values thus deduced are plotted. The doping dependence
of $\Delta^*$ is consistent with various spectroscopic data such
as tunneling and B$_{1g}$-symmetry Raman scattering \cite{Tacon}.
$\Delta^*\sim$ 30 meV for the $x$=0.15 sample is consistent with
the other ARPES results of single layer cuprates, too
\cite{KondoPRL,Terashima}. On the other hand, $\Delta_0$ remains
unchanged in going from $x$=0.15 to $x$=0.07, also the same as the
result of Bi2201 \cite{KondoNat}, indicating a behavior contrasted
with $\Delta^*$.

\begin{figure}
\begin{center}
\includegraphics[width=16cm]{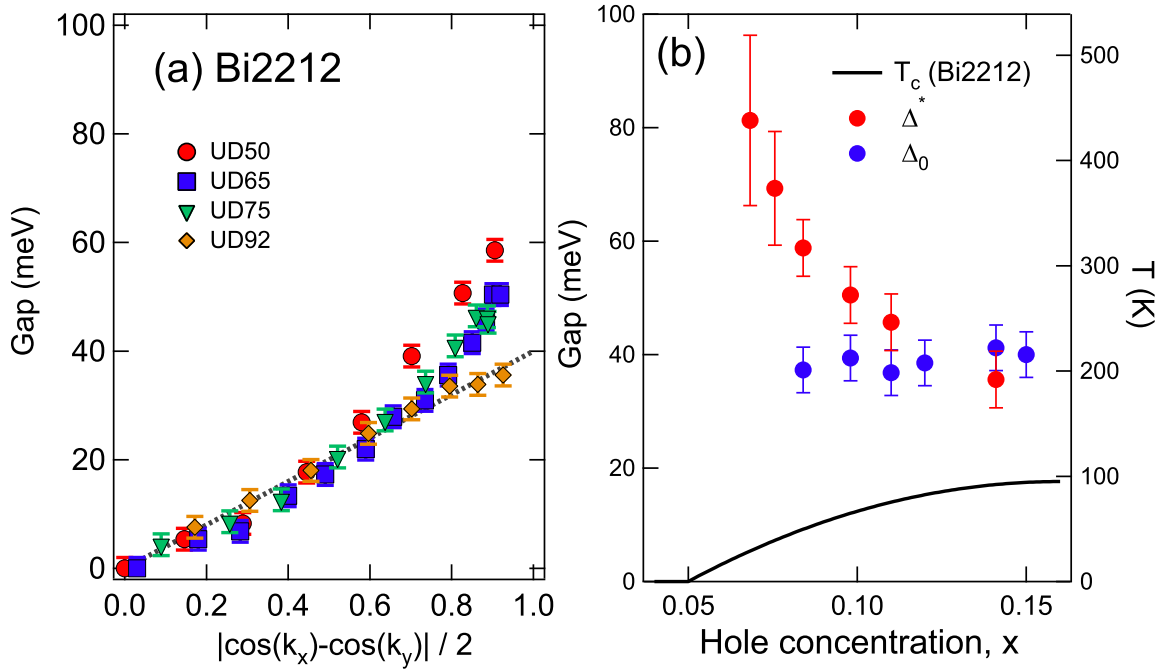}
\end{center}
\caption{(Color online) Energy gaps for underdoped Bi2212. (a)
Momentum dependence of the energy gap measured at 10 K
\cite{Tanaka, Lee}. A deviation from a simple $d$-wave form
(dashed line) is observed near the antinode in underdoped samples
with $T_c < $ 92K. (b) Doping dependences of the nodal gap
$\Delta_0$ and the anti-nodal gap $\Delta^*$. The $T_c$ dome of
Bi2212 is also plotted.} \label{Tanaka2}
\end{figure}

%Bi2212
Next, let us look at the energy gaps of bi-layer materials which
have much higher $T_c$'s than the single layer materials. Energy
gaps at 10K observed by ARPES for underdoped Bi2212 are presented
in Fig. \ref{Tanaka2} \cite{Tanaka, Lee}. As shown in Fig.
\ref{Tanaka2}(a), deviation from a simple $d$-wave form, becomes
prominent with decreasing hole concentration, which is the same
trend as the single layer cuprates as described above. On the
other hand, one can find that the gap function of slightly
underdoped Bi2212 with $T_c$= 92 K does not show a clear deviation
from the simple $d$-wave form. This is contrasted with the case of
the single layer cuprates, where the deviation from the simple
$d$-wave is prominent not only in the underdoped but also in the
optimally doped samples (Fig. \ref{Yoshida2}). This difference
stems from the much larger size of the near-nodal gap $\Delta_0$
in Bi2212 than those of the single layer cuprates, making the
condition $\Delta^* > \Delta_0$ difficult to study. Another
interesting point in Fig.\ref{Tanaka2}(a) is that the gap
functions for various doping levels near the nodal direction are
almost the same irrespective of the different $T_c$'s. In the same
manner as in Fig. \ref{Yoshida2}, the extracted $\Delta^*$ and
$\Delta_0$ values of Bi2212 are plotted as functions of doping in
Fig. \ref{Tanaka2}(b). $\Delta_0$ is nearly independent of the
hole concentration from the optimal to underdoped regions, in
contrast with the anti-nodal gap $\Delta^*$ which increases with
decreasing doping. Thus, not only $\Delta^*$ but also $\Delta_0$
do not trace the superconducting dome and, therefore, the
relationship between the gaps and the $T_c$ cannot be understood
from the simple mean-field picture.

\begin{figure}
\begin{center}
\includegraphics[width=16cm]{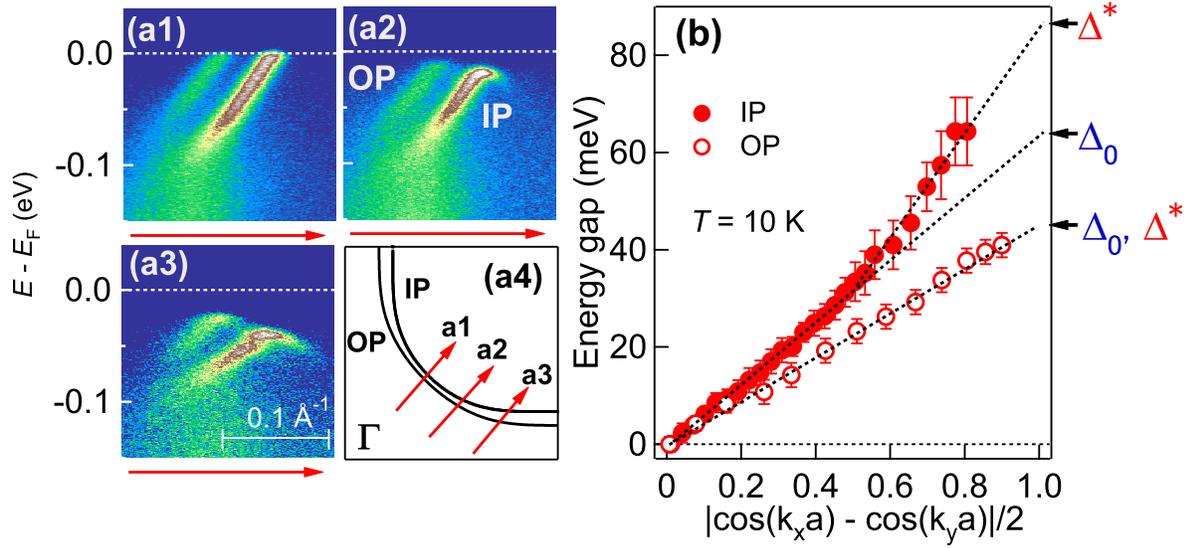}
\end{center}
\caption{(Color online) Energy gap in the superconducting state
($T$=10K) of optimally doped Bi2223 \cite{Ideta}. (a1)-(a3)
Energy-momentum intensity plots of the cuts from the nodal to
off-nodal regions for the outer and inner CuO$_2$ plane (OP and
IP) bands. The corresponding cuts are shown in panel (a4). These
data were taken with a photon energy of $h\nu$= 11.95 eV. (b)
Momentum dependences of the energy gaps for the OP and IP bands.
The definition of the nodal gap $\Delta_0$ and the antinodal gap
$\Delta^*$ is shown.} \label{Ideta1}
\end{figure}

%Bi2223
In the case of the tri-layer cuprate
Bi$_2$Sr$_2$Ca$_2$Cu$_3$O$_{10+\delta}$ (Bi2223), which has the
highest $T_{c,max}$( =110 K) among the Bi-family cuprates, energy
bands and Fermi surfaces originating from the outer and inner
CuO$_2$ planes (OP and IP) have been observed by Ideta \textit{et
al.} \cite{Ideta}. The hole concentration for the OP and IP bands
deduced from the Fermi surface areas are 23\% and 7\%,
respectively. Hence, the average hole concentration is 18\%, close
to the optimal doping concentration $\sim$16 \%. One may think
that the deduced hole concentrations are influenced by the
hybridization between the orbitals in neighboring CuO$_2$ layers.
However, the inter-layer hopping is small because the two OP bands
do not show clear splitting. Hence, the effects of the
hybridization on the estimated values would be negligible. In
Figs. \ref{Ideta1}(a1)-(a3), we show the dispersions of the OP and
IP bands in the superconducting state from the nodal to off-nodal
cuts \cite{Ideta}. The gap energies for both bands are very
different, as in the case of other multi-layer cuprate
Ba$_2$Ca$_3$Cu$_4$O$_8$F$_2$ (F0234) with $n$ = 4 \cite{Chen}. The
momentum dependence of the gap magnitude for OP is almost a simple
$d$-wave, $\Delta_0 |\cos(k_xa)-\cos(k_ya)|/2$ with $\Delta_0\sim$
43 meV, as shown by a straight line in Fig. \ref{Ideta1}(b). The
earlier ARPES results of Bi2223, where no clear band splitting has
been observed, give the gap value of $\sim$ 40 meV, possibly
reflecting the gap of the OP band \cite{Sato1,Sato2,Matsui}. On
the other hand, the gap for the IP band strongly deviates from the
simple $d$-wave around the anti-node $\sim$ ($\pi$, 0). The gap
size is therefore characterized by the two parameters
$\Delta_0\sim$ 60 meV around the node and $\Delta^*\sim$ 80 meV in
the aniti-nodal region. Because the deviation of the gap
anisotropy from the simple $d$-wave is prominent in the underdoped
cuprates, the observed gap anisotropy of the OP and IP are
consistent with the doping levels of the OP and IP estimated from
the FS areas. These energy gaps are much larger than those for the
same doping level of the double-layer cuprates, which leads to the
higher $T_c$ of Bi2223. Possible origins of the large
superconducting gaps for the OP and IP is the minimal influence of
out-of-plane disorder \cite{Eisaki} and/or interlayer tunneling of
Cooper pairs between the OP and IP \cite{Chakravarty}.

\begin{figure}
\begin{center}
\includegraphics[width=16cm]{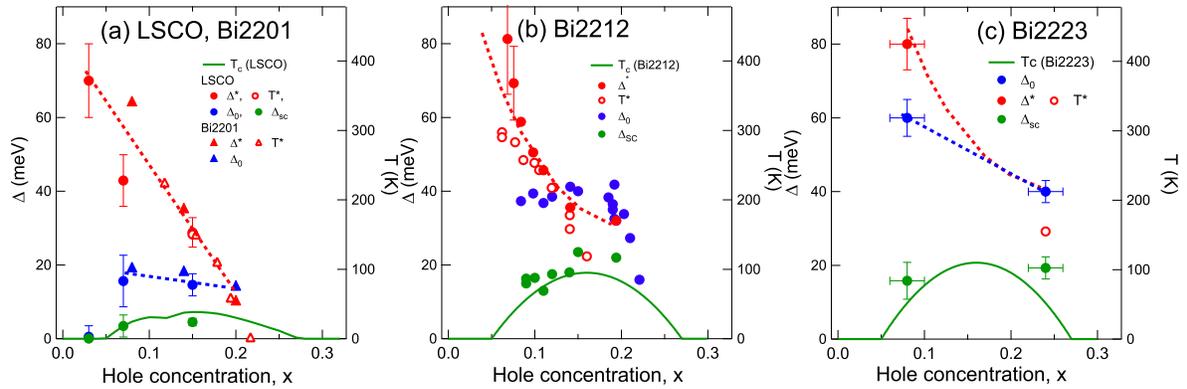}
\end{center}
\caption{(Color online) Doping dependences of the characteristic
energies ($\Delta^*, \Delta_0$) and temperatures ($T^*, T_c$) for
the single-layer cuprates (LSCO, Bi2201) (a), the double-layer
cuprates Bi2212 (b) and the tri-layer cuprates Bi2223 (c). Gap
energies $\Delta$ and temperatures $T$ have been scaled as
$2\Delta=4.3k_BT$ in these plots. Parameter values have been taken
from the ARPES studies of LSCO \cite{YoshidaTwoGap,Terashima},
Bi2201 \cite{KondoNat}, Bi2212
\cite{Tanaka,Lee,Campuzano,Ding,Feng} and Bi2223
\cite{Ideta,Sato1}. $T^*$ for Bi2201 and Bi2212 have been taken
from the NMR \cite{Zheng} and the transport studies
\cite{Konstantinovic}, respectively. $\Delta_{sc}$ defined in the
text approximately follows the $T_c$ dome. Dashed curves are guide
to the eye for $\Delta^*$ and $\Delta_0$.} \label{DopingGap}
\end{figure}

%Comparison with other cuprates
Now, let us compare the two gap energy scales for various kinds of
high-$T_c$ cuprates. In Fig. \ref{DopingGap}, the doping
dependences of $\Delta^*$ and $\Delta_0$ for the single-layer,
double-layer and tri-layer cuprates are plotted. Interestingly,
the doping dependence of the $\Delta^*$ of all these samples
approximately scale with the doping dependence of $T^*$ estimated
using various experimental techniques including ARPES as
$2\Delta^*/k_BT^*\simeq4$. Furthermore, in the underdoped region,
the doping dependences of the value of $\Delta^*$ and $T^*$ are
similar for all the systems irrespective of the very different
$T_{c,max}$ \cite{YoshidaTwoGap}. Judging from the underdoped
data, one can infer that $\Delta^*$ is a universal property of a
single CuO$_2$ plane and is not strongly affected by interaction
between layers.

%interpretation for Delta*
One possible explanation for the material independence of
$\Delta^*$ is that its magnitude is determined by the exchange
interaction $J$, since $J$ is almost material independent. A
pseudogap originating from antiferromagnetic spin fluctuations
\cite{Prelovsek, Kamimura} or RVB-type spin singlet formation
\cite{fukuyama} has its origin in $J$. Indeed, a phenomenological
model where the pseudogap $\Delta^*$ comes from the RVB gap
explains the two characteristic gap energy scales
\cite{Yang1,Yang2}. We should also note that, as discussed in the
next section, the behavior of the anti-nodal gap $\Delta^*$ is
consistent with a density-wave order model \cite{HashimotoNatPhys}
and may be relative to the nanoscale inhomogeneity observed by STM
\cite{Hanaguri}.

%interpretation for Delta0
 In contrast to the nearly material-independent $\Delta^*$, the nodal $d$-wave order parameter $\Delta_0$
of the optimally doped Bi2212 is twice as large as those of LSCO
and Bi2201. Furthermore, Bi2223 has even larger $\Delta_0$'s than
Bi2212 as shown in Fig.\ref{DopingGap}(c). These material
dependence roughly follows the magnitude of maximum $T_c$. The
strong material dependences of $\Delta_0$ imply that $\Delta_0$ is
not only a property of a single CuO$_2$ plane but also a property
influenced by the apical oxygens, the block layers and/or the
neighboring CuO$_2$ planes. Namely, the number of CuO$_2$ layers
and/or the distance of the apical oxygen atoms (in the block
layers) from the CuO$_2$ plane might be important factors for
determining the $\Delta_0$ and hence $T_{c,max}$. Influence from
outside the CuO$_2$ plane has been modelled using the
distant-neighbor hopping parameters, $t^\prime$ and
$t^{\prime\prime}$ \cite{Pavarini}, which are affected by the
$p_z$ orbital of the apical oxygen, the nearly filled Cu
$3d_{z^2}$ orbital, and the position of the empty Cu 4$s$ orbital.

\section{Temperature Dependence of the Energy Gaps and the Fermi Arc}
In the preceding section, we have demonstrated distinct doping
dependence of the near-nodal and near-antinodal gaps, as well as a
ubiquitous deviation from the simple $d$-wave form in underdoped
samples. This is because the superconducting gap and the pseudogap
may dominate in the nodal and antinodal regions, respectively. The
different natures of the pseudogap and the superconductivity are
also evident in their material dependences. In this section, we
further emphasize their distinct behaviors via temperature
dependent studies in different momentum regions. We discuss the
interplay between the pseudogap and the superconductivity, and
possible microscopic origins of the pseudogap.

\begin{figure}
\begin{center}
\includegraphics[width=12cm]{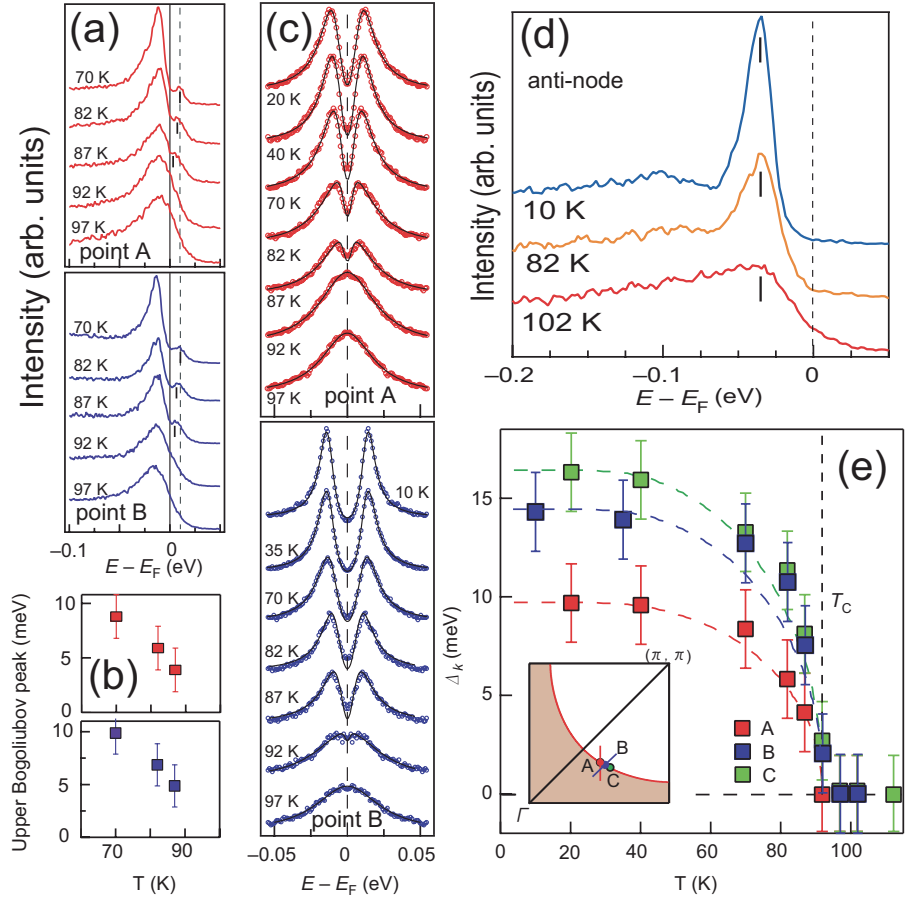}
\end{center}
\caption{(Color online) Temperature dependence of the gap in
Bi2212 \cite{Lee} (a) Temperature dependences of the raw spectra
at different $k_F$'s near the node across $T_c$ in underdoped
Bi2212 ($T_c$ = 92K, $p\sim$0.14) at point A (upper panel) and B
(lower panel). The upper Bogoliubov peaks are indicated by bars.
(b) Temperature dependence of the energy position of upper
Bogoliubov peaks. (c) Symmetrized EDCs across $T_c$ at point A
(upper panel) and B(lower penal). (d) Temperature dependence of
the EDC at the antinode on the Brillouin zone boundary across
$T_c$. Peak positions are indicated by bars. (e) Fitted gap size
as a function of temperature at different near-nodal momenta,
showing the BCS-like gap closing.} \label{TempDep1}
\end{figure}

Figure \ref{TempDep1}(a) shows the temperature dependence of the
gap in Bi2212 reported by Lee \textit{et al} \cite{Lee}. At low
temperatures, the symmetrized EDC's [panel (c)] on the Fermi
surface near the node show two peaks above and below $E_F$. They
shift towards each other with increasing temperature, indicating a
gradual closing of the gap, and eventually merge into a single
peak at $E_F$ above $T_c$, indicating the disappearance of the
gap. Correspondingly, the peak above $E_F$ in the raw spectra
[panel (a)], which can be understood as the upper branch of the
Bogoliubov quasi-particle, moves closer to $E_F$ with temperature
and disappears above $T_c$ [panels (a) and (b)]. Note that the
peaks abruptly disappear above $T_c$, which cannot be explained by
trivial thermal broadening. In panel (e), the temperature
dependences of the gap size for several cuts evaluated from the
symmetrized EDC's are summarized. The near-nodal gap closes at
$\sim T_c$, reminiscent of the BCS theory, although the exact
functional form may be different. This indicates that the gap near
the node is primarily of the superconductivity origin.

The lowest binding energy peak in the antinodal region, on the
other hand, does not show a strong temperature dependence across
$T_c$, from the underdoped to the overdoped samples, as shown in
Fig. \ref{TempDep1}(d). Interestingly, without closing the gap,
the sharp superconducting QP peak almost disappears above $T_c$
\cite{Lee,FengSci}. The intensity of the QP peak below $T_c$
scales with the superfluid density and the condensation energy
\cite{FengSci}. This suggests that the temperature dependences of
the gap magnitude and the spectral intensity in the antinodal
region does not follow the weak-coupling BCS theory. Therefore, it
is difficult to explain these two different types of temperature
dependences in the different momentum regions of a single Fermi
surface using a single order parameter. The existence of multiple
order parameters is hence implied, and this further emphasizes the
two-gap behavior and the nodal-antinodal dichotomy. Here, it
should be noted that the dichotomous temperature dependence is not
a sharp dichotomy in momentum space, but rather a crossover from
the temperature dependence near the node to the temperature
independence near the antinode. In fact, the temperature
dependence in the intermediate momentum region shows intermediate
behaviors: the gap becomes smaller but does not close completely
above $T_c$. Moreover, at the lowest temperature even the slightly
underdoped Bi2212 (UD92K) shows nearly the simple $d$-wave form,
despite that the temperature dependences near the node and
antinode are completely different as described above. Here, it
should be noted that we also observe a superconducting QP peak
even in the anti-nodal region well below $T_c$\cite{Vishik, Lee}.
This suggests that contribution to the condensation energy from
the anti-nodal electronic states is finite even though it is
weakened by the pseudogap opening. It seems natural to consider
that the superconducting gap near the antinode is strongly
distorted by the pseudogap, leading to the deviation of the gap
function and the coherent QP intensity from the simple $d$-wave
functional form.

Because of the nodal-antinodal dichotomy in the temperature
dependence of the gap, as shown in Fig.\ref{TempDep1}, slightly
above $T_c$, only a portion of the Fermi surface is recovered near
the node. We refer to this ungapped portion of the Fermi surface
above $T_c$ as the ``Fermi arc". Here, we define the Fermi arc by
the momentum region where symmetrized EDCs at $k_F$ have a single
peak at $E_F$ at each temperature above $T_c$. It thus looks as if
the original hole-like Fermi surface is truncated into four pieces
in the first Brillouin zone, although there are still debates
whether the ungapped portion is actually part of the original
Fermi surface or part of small hole pockets \cite{MengNat,Yang,
Kamimura}. The Fermi arc length seems proportional to the hole
concentration and is shown to increase with temperature
\cite{Kanigel}. As shown below, the Fermi arc length slightly
above $T_c$ will be used in the analysis between energy gap and
$T_c$.

\begin{figure}
\begin{center}
\includegraphics[width=16cm]{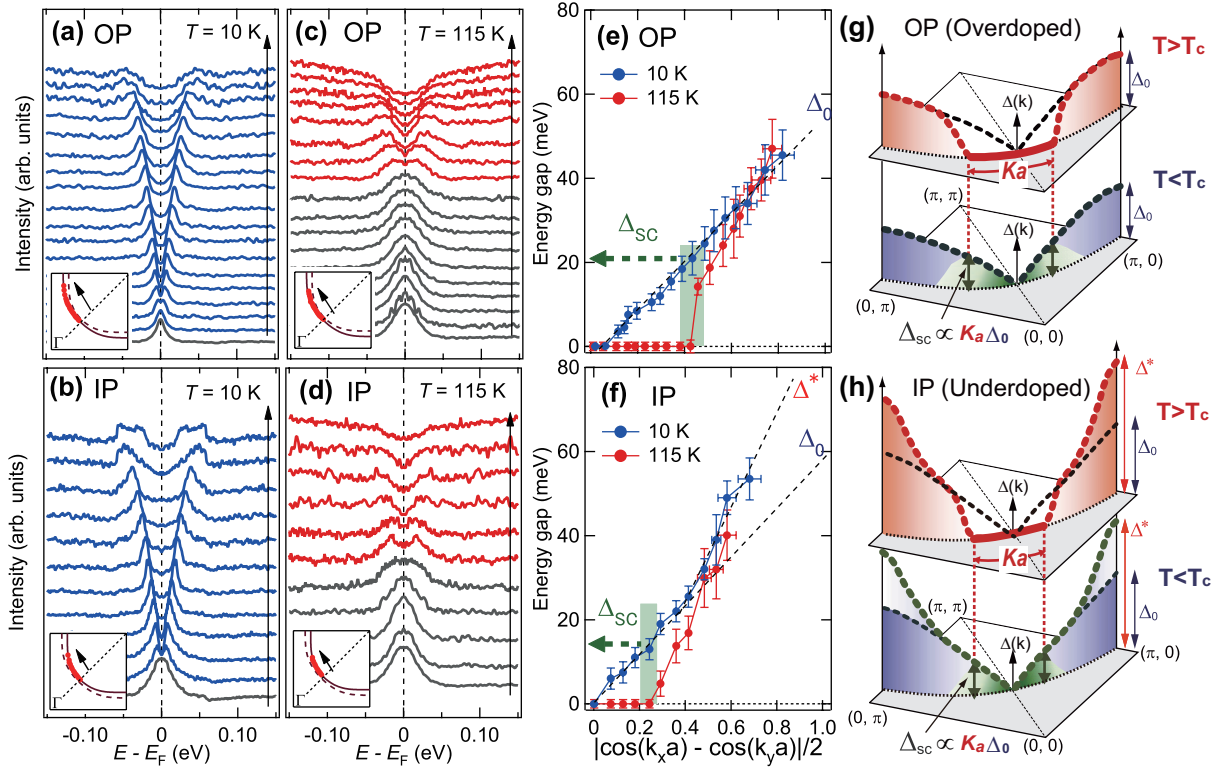}
\end{center}
\caption{(Color online) Momentum dependence of the energy gap
along the Fermi surface at $T \ll T_c$ and $T>T_c$ in optimally
doped Bi2223 ($T_c$ = 110 K) \cite{IdetaArc}. (a) Symmetrized EDCs
along the outer plane (overdoped) Fermi surface at 10 K. (b)
Symmetrized EDCs along the inner plane (underdoped) Fermi surface
at 10 K. (c)(d) Symmetrized EDCs above $T_c$ ($T =$ 115 K). (e)(f)
Fitted gap size below and above $T_c$ along the Fermi surfaces for
the outer and inner planes (OP and IP), respectively. (g)(h)
Schematic drawings of the evolution of the gap with temperature
for the outer and inner planes, respectively. The Fermi arc length
($K_a$), $\Delta_0$, $\Delta^*$ and effective superconducting gap
$\Delta_{sc}$ are defined.} \label{TempDep2}
\end{figure}

In Fig. \ref{TempDep2}, we show  another case of the temperature
dependence of the gap, namely, tri-layer cuprates Bi2223 reported
by Ideta \textit{et al}\cite{IdetaArc}. As in the case of Bi2212,
the gap near the node has the simple $d$-wave form and closes just
above $T_c$ to form a Fermi arc, again suggesting the major
contribution of the nodal region to the superconductivity. The gap
function of the (underdoped) IP clearly shows the two-gap behavior
well below $T_c$, and the Fermi arc length slightly above $T_c$ is
shorter than that for the OP, reflecting the lower hole
concentration of the IP. The gap in the antinodal region of the
(overdoped) OP remains open already above $T_c$ with little
temperature dependence, although the gap function well below $T_c$
follows the $d$-wave form. A QP peak exists on the entire Fermi
surface similar to the Bi2212 case. Similar behaviors have also
been reported in the single-layer cuprate families \cite{Meng,
KondoPRL,KondoNat,Ma}. These observation lead the conclusion that
the nodal-antinodal dichotomy and the existence of the Fermi arc
are the universal features of the cuprate. The almost BCS-like
temperature dependence of the gap on the Fermi arc suggests that
the Fermi arc is closely related to the superconductivity.

\begin{figure}
\begin{center}
\includegraphics[width=16cm]{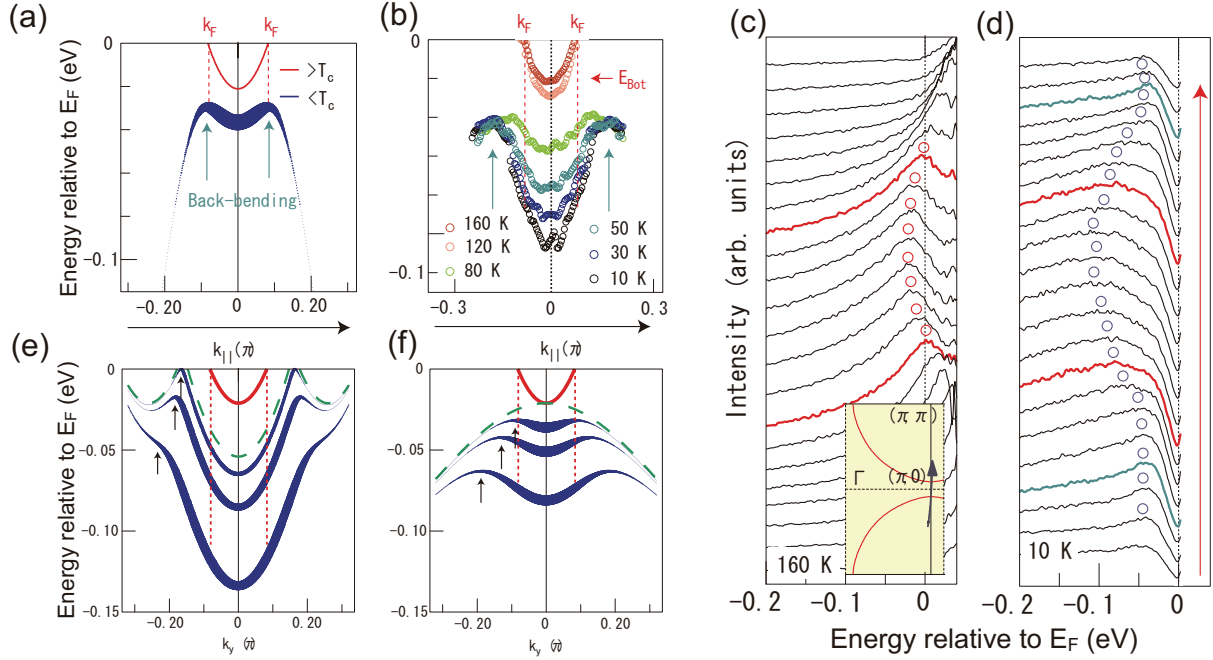}
\end{center}
\caption{(Color online) Particle-hole symmetry breaking in the
antinodal pseudogapped dispersion of the single layer Bi2201. (a)
Simulated dispersion for $d$-wave homogeneous superconductivity
with order parameter $V$ = 30 meV along
$(-\pi,\pi)-(\pi,0)-(\pi,\pi)$. Note that the back-bending
momentum is aligned to $k_F$. (b) Experimentally determined
temperature evolution of the dispersion across $T^* \sim$ 125 K.
(c) EDCs in the ungapped state above $T^*$. Inset shows the
measured cut. (d) EDCs at 10 K. Red EDCs are for $k_F$. The
dispersion shows no anomaly at $k_F$. (e)(f) Simulated dispersion
for Two long-range orders: (e) incommensurate checkerboard
density-wave order of orthogonal wave vectors, [0.26$\pi$, 0], [0,
0.26$\pi$] and (f) commensurate [$\pi,\pi$] density-wave order
with order parameter $V$ = 15, 30 and 60 meV (from top to bottom).
Back-bendings are indicated by arrows.
 } \label{TempDep3}
\end{figure}

Focusing on the detailed temperature dependence of the antinodal
pseudogap of the single-layer Bi2201, Hashimoto \textit{et al}.
have shown that the antinodal gap is not caused by
superconductivity but it is more consistent with a short-range
density-wave order by studying the gap energy and the entire band
dispersion \cite{HashimotoNatPhys, RHHScience}. Since the $T_c$
($\sim$ 34 K) of this system is low and the difference between
$T^*$ ($\sim$ 125 K) and $T_c$ is relatively large, one can study
the pseudogap state between $T_c$ and $T^*$ in great detail. As
one can see from Figs. \ref{TempDep3}(b) and (c), above $T^*$, the
ARPES spectra along an antinodal cut [inset of Fig.
\ref{TempDep3}(c)] show a parabolic dispersion crossing $E_F$.
This is strikingly simple, like the dispersion in a simple metal,
and we define $k_F$ unambiguously from the Fermi crossing points.
Note that, since the superconducting peak is very weak in this
compound, the pseudogap dispersion can be studied below $T_c$. The
pseudogap opens at $T<T^*$, and spectra at these lower
temperatures are paradoxically broader than at higher temperature.
We track the dispersion by the peak positions of EDCs at different
$k$ points. In the 10 K spectra displayed in Fig.
\ref{TempDep3}(d), no ``back-bending" is observed at $k_F$ defined
above $T^*$. Instead, the dispersion bends back at momenta
markedly away from the $k_F$. Very importantly, this misalignment
of the back-bending momentum and $k_F$ cannot be explained by the
opening of a simple superconducting gap, which requires a gap with
particle-hole symmetry. As shown in Fig. \ref{TempDep3}(a), one
always expects the alignment of $k_F$ and the ``back-bending"
momentum of the dispersion in the superconducting state. Because
of this strong constraint, it can be concluded that the observed
behavior below $T^*$ is different from the expectation for the
superconducting state, suggesting that the transition from the
true normal state above $T^*$ to the pseudogap state has a
different origin from Cooper pairing. The smooth temperature
evolution upon cooling from the true normal state
\cite{HashimotoNatPhys} suggests a direct connection of the
broken-symmetry state below $T^*$ with the pseudogap opening.
Also, the temperature dependence of the antinodal gap in the ARPES
spectra coincides with the results of polar Kerr rotation and
time-resolved reflectivity \cite{RHHScience}. This particle-hole
asymmetric gap may be explained by some sort of density wave order
with short correlation length. It is found that the simulations of
both the checkerboard density-wave order of orthogonal wave
vectors and commensurate ($\pi,\pi$) density-wave order can
qualitatively reproduce the misalignment of the back-bending
momentum and $k_F$ and the stronger dispersion at lower
temperatures. The anomalous broadening upon pseudogap opening
could be understood if finite correlation length of the density
wave order is considered \cite{HashimotoNatPhys, RHHScience}.
Intriguingly, the density wave order considered here could be
consistent with the momentum-integrated STM observations in real
space \cite{Wise}. Furthermore, broken spatial symmetry with
nematic order and its finite correlation length have been reported
in the pseudogap phase \cite{Lawler}. The observed nanoscale
inhomogeneity associated with local density-wave order is formally
consistent with the spatially-averaged observation by ARPES, and
supports that the pseudogap is a broken-symmetry state below
$T^*$. This pseudogap most likely competes with superconductivity,
leaving the form of the interaction between the pseudogap and
superconductivity as an open question.

%\section{Fermi arc picture}
%universal relationship
Now, we turn to the subject of scaling relationship related $T_c$
in the high-$T_c$ cuprates. The well known Uemura relationship
relates the superfluid density with $T_c$ \cite{Uemura} for
underdoped cuprates. Tallon \textit{et al.} \cite{Tallon} have
proposed a modified Uemura relation in which the value of
$T_c/\Delta^\prime$ plotted as a function of superfluid density,
where $\Delta^\prime$ is the maximum spectral gap obtained from
the specific heat and Raman studies. Since an energy gap
presumably due to competing order is open already above $T_c$, one
may infer that the pseudogap weakens contribution to the
condensation energy from the anti-nodal region as indicated by the
superconducting peak ratio \cite{FengSci}. Then, we propose
another relationship between the Fermi arc and $T_c$. Here, we
define the ``effective" superconducting gap $\Delta_{sc}$, by the
gap at the end point of the Fermi arc slightly above $T_c$,
[$\propto$(Fermi arc length) $\times \Delta_0$]. Then,
$\Delta_{sc}$ rather than $\Delta_0$ would be more directly
related to $T_c$ \cite{Oda}. Note that this is a $T=T_c$ scaling
relation rather than a $T=0$ scaling relation like the Uemura
relation.

\begin{figure}
\begin{center}
\includegraphics[width=16cm]{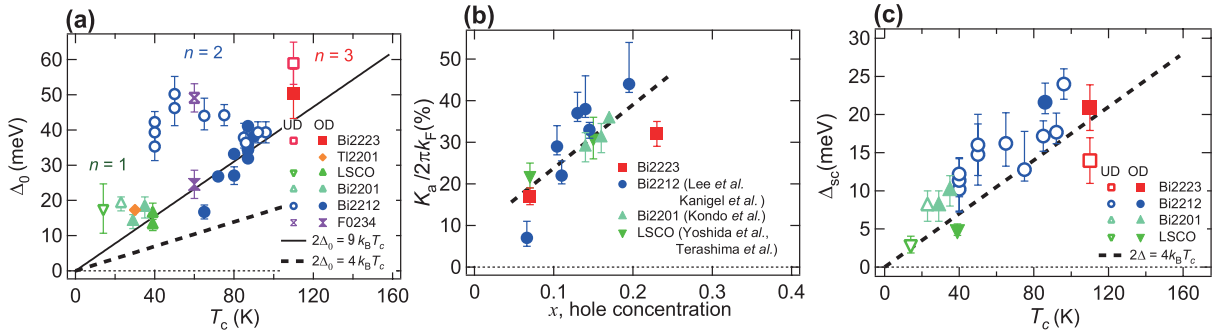}
\end{center}
\caption{(Color online) Relationship between the nodal
superconducting gap $\Delta_0$, the Fermi arc length $K_a$, the
superconducting gap at the edge of the Fermi arc $\Delta_{sc}$ and
$T_c$ for various high-$T_c$ cuprates. (a) $\Delta_0$ as a
function of $T_c$. Data are taken from ARPES results for LSCO,
Bi2201, Bi2212, Bi2223 in Fig.\ref{DopingGap} and F0234
\cite{Chen}. (b) $K_a$ relative to that of the full Fermi surface
2$\pi k_F$ as a function of hole concentrations. Dashed line is a
guide to the eye. (c) $\Delta_{sc}$ as a function of $T_c$. The
dashed line indicates 2$\Delta$=4 $k_BT_c$, which is close to the
weak coupling $d$-wave BCS relationship.} \label{universal}
\end{figure}

%Delta0 vs Tc
In Fig. \ref{universal}(a), we plot the nodal superconducting gap
$\Delta_0$ measured by ARPES (Fig. \ref{DopingGap})
\textit{versus} $T_c$ for various high-$T_c$ cuprates. For
optimally-doped to overdoped samples, the experimental data follow
the relationship $2\Delta_0\sim9k_BT_c$, reminiscent of a strong
coupling formula of $d$-wave superconductivity, and in the
underdoped region, the plot becomes $2\Delta_0 \gg 9k_BT_c$,
deviating from the linear relationship between $\Delta_0$ and
$T_c$. Next, we apply our scaling analysis with the Fermi arc
length $K_a$. In Fig. \ref{universal}(b), $K_a$ for LSCO
\cite{YoshidaTwoGap,Terashima}, Bi2201 \cite{KondoPRL}, Bi2212
\cite{Kanigel,Lee}, and the OP and IP of Bi2223 \cite{IdetaArc}
are plotted as a function of doped hole concentration $x$. One
finds that the $K_a$ values of the various high-$T_c$ cuprates
approximately fall on a single line. Here, we should note that the
magnitude of the antinodal gap $\Delta^*$ in the underdoped
region, which is nearly material independent, may have an inverse
correlation with $K_a$, which may cause a material independence of
$K_a$. The evolution of $K_a$ with $x$ implies increase of holes
in the normal state. Tanner \textit{et al.} \cite{Tanner} pointed
out that the number of superconducting electrons $n_s$ is
empirically proportional to the number of normal state carrier
$n_n$: $n_s \sim 0.2n_n$. Therefore, the Fermi arc length also
scales with superfluid density $\rho_s = n_s/m^*$ \cite{KondoNat},
where $m^*$ is the carrier effective mass.

In Fig. \ref{universal}(c), we have plotted the ``effective"
superconducting gap $\Delta_{sc} = \Delta_0 \sin(K_a/k_F)$ against
$T_c$ for various high-$T_c$ cuprates. These data approximately
fall on the straight dotted line ($2\Delta = 4k_BT_c$) which is
close to the $d$-wave BCS gap ratio $2\Delta_\mathrm{BCS} =
4.3k_BT_c$ \cite{Maki}. It should be noted that this relationship
seems to hold even in the overdoped region. Hence, the
``effective" superconducting gap $\Delta_{sc} \propto K_a\Delta_0$
scales with $T_c$ better than $\Delta_0$ does. This relationship
is reminiscent of the relationship $T_c \propto x\Delta_0$
proposed by Lee and Wen \cite{LeeWen} and Oda \textit{et al.}
\cite{Oda}. The closer relationship between $\Delta_{sc}$ and
$T_c$ than that between $\Delta_0$ and $T_c$ implies that
contribution to the condensation of superfluid from the anti-nodal
region may be reduced due to the gap opening above $T_c$. While
the relationship between $\Delta_{sc}$ and $T_c$ is close to the
weak-coupling BCS formula $2\Delta_{BCS} \simeq 4.3k_BT_c$ [Fig.
\ref{universal}(c)], $\Delta_0$ satisfies the strong-coupling
relationship $2\Delta_0 \simeq 9k_BT_c$ [Fig. \ref{universal}(a)]
from the optimum to the overdoped regions. When the gap has a
simple $d$-wave form in the overdoped samples, $\Delta_0$ may be
comparable or larger than the magnitude of the gap from competing
order. How the competing order fades away with doping in the
overdoped region is an open question to be addressed in future
studies.

\section{Conclusion}
 We have presented an overview of the two-gap behaviors in the
high-$T_c$ cuprates observed by ARPES and described the doping,
material, and temperature dependences of the energy gaps. Two
distinct energy scales $\Delta^*$ and $\Delta_0$ have been
identified to describe the two-gap energy structure where their
momentum dependence sometime deviates from the simple $d$-wave.
The observed energy gaps show dichotomy between the near nodal
(characterized by $\Delta_0$) and anti-nodal direction
(characterized by $\Delta^*$) regions. While $\Delta^*$ at the
same doping level is nearly material independent, $\Delta_0$ shows
a strong material dependence reflecting the magnitude of
$T_{c,\mathrm{max}}$. As for the temperature dependence, the gap
near the nodal direction closes at $T_c$. In contrast, the
anti-nodal gap does not close at $T_c$ and its spectral features
cannot be explained by conventional superconductivity, but
consistent with a short-range density-wave picture
\cite{HashimotoNatPhys}. Therefore, the origin of $\Delta^*$ is
likely to be a density-wave order competing with the
superconductivity, although finite contribution of the
pseudogapped region to the superconductivity may also exist. From
$\Delta_0$ and the Fermi arc length, $K_a$, we define an
"effective superconducting gap" $\Delta_{sc}$, which scales with
$T_c$ through $2\Delta_{sc} \simeq 4k_BT_c$, though this
relationship does not imply that superconductivity is absent away
from the Fermi arc below $T_c$. The results give a reasonable
scaling with $T_c$ for different materials. To fully elucidate the
two-gap behavior, both phenomena, the competing order and the
superconducting fluctuations, should be properly taken into
account. This issue should be clarified in future studies of the
electronic structure by ARPES and STM.

\section*{Acknowledgement}
This work was supported by a Grant-in-Aid for Young Scientist B
(22740221), the A3 Foresight Program from the Japan Society for
the Promotion of Science, and by DOE Office of Basic Energy
Science, Division of Materials Science (Contract number
DE-FG03-01ER45929-A001 and DE-AC02-76SF00515)

\bibliography{PG_ARPES}

\end{document}